\documentclass[twocolumn,showpacs,preprintnumbers,amsmath,amssymb]{revtex4}
\usepackage{amsmath,amssymb,ams,bbm,epsfig}
\newcommand{\Uop}{{\hat U}}

\newcommand{\Dt}{\frac {\tau}{L}}
\newcommand{\uu}{\tilde u}

\newcommand{\ii}{\mbox{i}}
\newcommand{\iis}{\mbox{\tiny i}}

\begin{document}

\title{Delocalized and Resonant Quantum Transport 
in Nonlinear Generalizations of the Kicked Rotor Model}
\author{Laura Rebuzzini,$^{1}$ Sandro Wimberger,$^2$ and Roberto Artuso$^{1,3,4}$}
\affiliation{$^1$Center for nonlinear and complex systems and Dipartimento di Fisica e Matematica, \\
Universit\`a dell'Insubria, Via Valleggio 11, 22100 Como, Italy \\
$^2$Dipartimento di Fisica E. Fermi, Universit\`{a} degli Studi di Pisa, 
Via Buonarroti 2, 56127 Pisa, Italy\\
$^3$Istituto Nazionale per la Fisica della Materia,
Unit\`{a} di Como, Via Valleggio 11, 22100 Como, Italy \\
$^4$Istituto Nazionale di Fisica Nucleare, Sezione di Milano,
Via Celoria 16, 20133 Milano, Italy}

\begin{abstract}
We analyze the effects of a nonlinear cubic perturbation on the $\delta$-Kicked
Rotor. We consider two different models, in which the
nonlinear term acts either in the position or in the momentum
representation. We numerically investigate the modifications induced by the
nonlinearity in the quantum transport in both localized and resonant regimes
and a comparison between the results for the two models is presented. 
Analyzing the momentum distributions and the increase of the mean
square momentum, we find that the quantum resonances asymptotically
are very stable with respect to the nonlinear perturbation of the rotor's 
phase evolution. For an intermittent time regime,
the nonlinearity even enhances the resonant quantum transport, 
leading to superballistic motion.
\end{abstract}

\date{\today}
\maketitle

%%%%%%%%%%%%%%%%%%%%%%%%%%%%%%%%%%%%%%%%%%%%%%%%%%%%%%%%%%%%%%%%%%%%%%%%%%%%%%%

%%%%%%%%%%%%%%%%%%%%%%%%%%%%%%%%%%%%%%%%%%%%%%%%%%%%%%%%%%%%%%%%%%%%%%%%%%%%%%%
\section {Introduction}
Recent and ongoing experiments \cite{expintro} have started to investigate
the interplay between the many-body induced self-interaction in an ultracold
atomic gas and an external driving induced by time-dependent optical
potentials. The natural setup is to use a Bose-Einstein condensate of alkali
atoms, where the nonlinearity parameter can be tuned \cite{PS2002,DGPS}, 
and pulsed optical lattices can be used to impart momentum kicks on 
the atoms. For such a setup, the Gross-Pitaevskii (GP) equation \cite{PS2002,DGPS}
provides a good description of the system, as
long as the nonlinearity is not too large, as a study of the stability
of linearized excitations around the GP solution has shown
\cite{Zoller2000,Raizen2004}.

In this paper, we analyze the evolution of a cubic 
nonlinear Schr\"odinger equation, as present in the GP model,
under the perturbation of time-periodic $\delta$-kicks
\begin{equation}
\label{sch}
\ii \frac {\partial\psi}{\partial t'}= \left[-\frac 12 
\frac {\partial^2}{\partial \vartheta^2} - u|\psi |^2 
+k \cos(\vartheta ) \sum _{t=0}^{+\infty}\delta (t'-t\tau ) \right] \psi,
\end{equation}
where $\vartheta$ and $n=-\ii\frac {\partial}
{\partial\vartheta}$ are the  position and the conjugated momentum 
of the system; we chose units such that $\hbar =1$ and the motion is 
considered on a ring with periodic boundary conditions 
$\psi (\vartheta + 2\pi ) = \psi (\vartheta)$. 
The parameters $u$ and $k$ are the nonlinearity coupling 
and the kicking strength, respectively.

In atom optics experiments, the $\delta$-kicked rotor has been realized 
with an ensemble of laser-cooled, cold atoms \cite{expold}, 
or recently also with an ultracold Bose-Einstein condensate \cite{expintro},
periodically driven with a standing wave of laser light. With
the wave number of the laser $k_L$, 
the experimental variables are easily expressed in our units 
by noting that momentum is usually measured in two photon recoils
($2\hbar k_L$), and position in units of the inverse wave
number of the standing wave ($1/2k_L$). Hence, 
the scaled variables $\vartheta,n$ and the physical ones 
$\vartheta',p'$ are related by $\vartheta =2k_L\vartheta'$ and $n=
p'/2k_L\hbar$ \cite{darcy2004,WGF2003,SMOR}. Our choice of units makes 
all the relevant quantities (including the ones plotted in figures) 
dimensionless.

Owing to the periodicity of $\delta$-kick perturbation, the time $t$ 
is measured in number of periods $\tau$ and the evolution of 
the wave function of the system over a time interval $\tau$ is described by the operator 
$\Uop (\tau)$.

A cubic modification of linear Schr\"odinger dynamics for
the $\delta$-kicked rotor may be 
accomplished by two different models, both considered in the present paper. 
The correct way to approximate the evolution of
the nonlinear Schr\"odinger equation is to evaluate the nonlinear term
in the position representation \cite{BCPS1991}. In the following we will 
refer to this first model as model 1 (M1). \\
Since the Hamiltonian operator presents the time dependent nonlinear part 
$u|\psi |^2 \psi$, in the numerical integration of Eq.(\ref{sch}), the lowest 
order split method \cite{split} is used and $\Uop$
is approximated by the time ordered 
product of evolution operators (Trotter-Kato discretization \cite{RS1980}) 
on small time steps $\tau /L$ (with $L$ integer):
\begin{eqnarray}
\label{model1}
 \Uop ^{(1)}(\tau)&=& \hat K\hat R^{(1)}(\tau) \nonumber \\
&\approx&  e^{-\iis k\cos(\hat\vartheta)}\prod _{l=1}^{L} e^{-\iis
\Dt \frac {\hat n^2}{2}}\ e^{\iis u\Dt |\psi (\hat\vartheta,l\Dt )|^2}.
\end{eqnarray}
In the numerical simulations, we use a finite Fourier basis 
of dimension $N$: 
the discrete momentum eigenvalues lie on the lattice 
$p=(m-\frac N2)$ and the continuous angle variable is approximated by 
$\vartheta =\frac {2\pi}N(m-
1)$ with $m\in {\bf Z}, 1\leq m\leq N$. 
Shifting between the coordinate and momentum representations, in 
the evaluation of the operator $\hat R^{(1)}(\tau) $, requires
$2L$ Fast Fourier transforms of $N$-dimensional vectors for each kick. 
In order to get stable numerical 
results, the splitting interval $\tau /L$ has to be reduced 
when increasing the nonlinear coupling constant $u$; typical values 
of the number of steps per period 
range between $L=80000$ and $L=5000000$. Therefore, investigating 
either the effects of strong nonlinearities or the dynamics of 
the system over long times is computationally quite expensive with (M1).  

A second model (known as the Kicked Nonlinear Rotator and 
called in the following model 2 (M2)) was introduced in \cite{Shep1993}.
This model, being a much simpler variant of the Kicked Rotator 
Model (KR) \cite{CCFI79}, 
allows to perform faster and more efficient numerical computations. \\
The evolution operator over one period $\tau$ for (M2) is given by
 \begin{eqnarray}
\label{model2}
\Uop ^{(2)}(\tau) &= &\hat K\hat R^{(2)}(\tau) \nonumber \\
&\approx &e^{-\iis k\cos(\hat\vartheta)} e^{-\iis 
\tau\frac {{\hat n}^2}{2}} e^{\iis \uu \tau |\hat \psi _n|^2}\, 
\end{eqnarray}
where $\hat \psi _n$ indicates the $n$-th component of the wave 
function of the system in the momentum 
representation. The change in the phase of each component 
$\hat \psi _n$ of the state vector, introduced by the nonlinear 
term between two kicks, is proportional to the amplitude of the component. 
In (M1) instead, the phase acquired at each instant by the wave function 
involves all the Fourier 
components and the phase factor has as the $n$-th Fourier component
%(in the momentum representation) attached to the component $\hat \psi_n$ is
$\frac u{2\pi}\sum _m\hat \psi^* _{m+n}
\hat \psi _{m}$. The two models coincide 
only if the wave function of the system is a plane wave of fixed momentum;
in this case, the relation between 
the nonlinear coupling constants in (M1) and (M2) is $\uu=u/2\pi$.
%% saw
%%In the following calculations, the initial state is always chosen 
%%in the form of a plane wave of null momentum ($\hat \psi_n(0) =\delta (0)$).

Both models (M1) and (M2) are nonlinear generalizations of the KR 
and reduce to the KR in the limit $u\to 0$. 
Depending on the commensurability of the period $\tau$ 
of the $\delta$-kick perturbation with $4\pi$, the KR displays different 
regimes, deeply studied both theoretically 
\cite{CCFI79,chir,Izr1990} 
and experimentally \cite{expold,darcy2004}. 
The quantum resonant regime, corresponding to values of $\tau$ being rational 
multiples of $4\pi$, is characterized by a ballistic transport: the 
mean energy of the system grows according to a parabolic law
\cite{IzShep,Izr1990}. 
For generic irrational values of $\tau$, the average energy grows linearly 
in time only within a characteristic time (break time), 
after which dynamical localization sets in and the diffusion 
is suppressed \cite{chir}.  

In this paper, we analyze in detail how the presence of the 
nonlinearity affects 
the general properties of transport in regimes that correspond 
to the localized (Section II, which is essentially a warm up exercise in which
the results of \cite{Shep1993} are reproduced and new numerical results about 
prefactor scaling is presented) and resonant (Section III) 
one of the KR. We focus the attention on the 
growth exponent of the mean energy and on how the diffusion coefficient 
or the rate of ballistic transport depend on the strength 
of the nonlinear coupling constant $u$.
While in Sections (II) and (III) we are dealing with the evolution of an
initial state with fixed momentum, chosen at $n=0$, i.e. 
$\hat \psi_n(0) =\delta (0)$, Section
(IV) is devoted to the effects of a finite spread of initial conditions as
strongly suggested by state-of-the-art experiments using ultracold atoms
\cite{expintro,Pisa2001}.

The quantity we typically compute is the width of the momentum distribution 
of the system $\langle p^2(t)\rangle=\sum _{n=-\infty}^{+\infty}
n^2 |\hat\psi _n |^2$, which 
gives the spreading 
of the wave packet over unperturbed levels or, equivalently,
 - apart from a constant factor 2 -
the expectation value of the energy. 
The time-averaged spreading  
$P^2(T)=\frac 1T \sum _{t=1}^{T}\langle p^2(t)\rangle$ of the 
second moment has been frequently used (see {\em e.g.} \cite{AR2002}), 
as it preserves the exponent of the power law growth, 
while smoothing out oscillations.

As pointed out in \cite{Shep1993} the nonlinear shift is essential in determining
dynamical features (providing for instance a mechanism for delocalization of the
generic, irrational case), but when we deal with delocalized states it is typically
quite small ({\em e.g.} for (M2) the shift is proportional to $\hat{u}/\Delta n$, where
$\Delta n$ is the width of the distribution over unperturbed states). So in general we
may expect that, for moderate nonlinearities, the precise form of the shift does not
alter in an essential way the nature of asymptotic motion.

%Even if differences exist on small time scales, the two models (M1) and (M2)
%seem to share similar asymptotic behaviors.
%This can be explained as follows. The presence of the nonlinearity causes a 
%delocalization of the system and the consequent growth of the 
%momentum distribution $\Delta n$. 
%The nonlinear phase shift $\Phi (t)$ depends on the amplitude 
%$|\hat\psi_n|^2$, which, owing to the normalization condition, 
%decays in inverse ratio to the spreading of the wave packed, 
%$|\hat\psi_n|^2\sim 1/\Delta n$. 
%{\bf PLEASE COMMENT: 
%A similar equilibration of the wave function occurs in the position
%representation.} 
%Therefore the spreading in time of the distribution 
%produced by the nonlinearity leads to a decrease in the nonlinear 
%shift $\Phi (t)$. 

\section {Localized regime}

\begin{figure}
\centerline{\epsfxsize=8.5cm \epsfbox{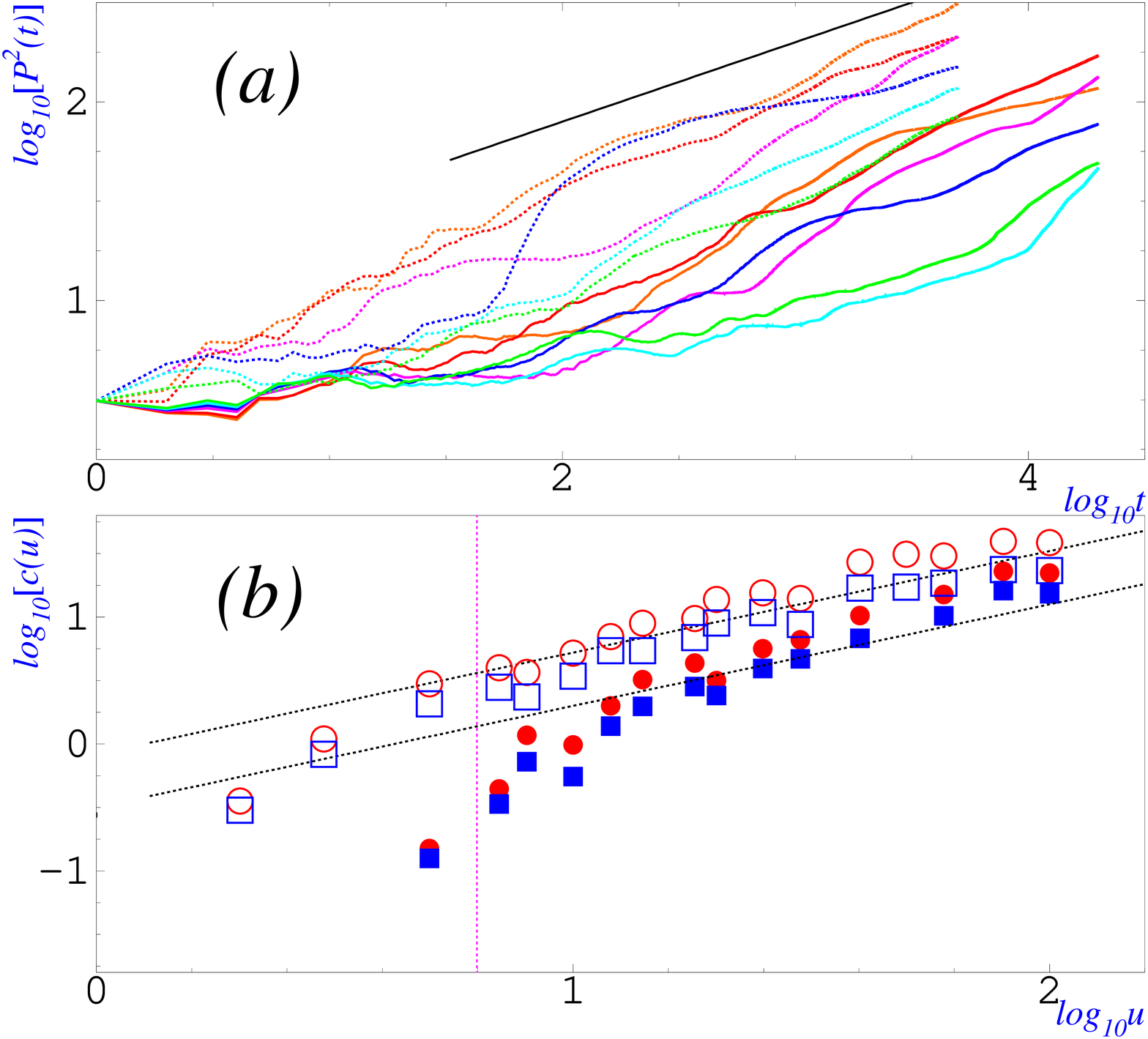}}
%\centerline{\epsfxsize=7cm \epsfbox{dipbeta-loc.eps}}
\caption{\small 
(Color online) (a) Bilogarthmic plot of the time averaged 
second moment vs time in the localized regime. Time is 
measured in number of periods. The dashed and full lines refer 
to the (M2), with $N=2^{17}$, and (M1), with $N=2^9$ and $L=80000$, 
respectively. 
Values of 
$u 
= 8,\ 10,\ 12,\ 14,\ 16,\ 20$ are considered; generally higher 
nonlinearity values yield bigger spreading.
The dashed line has the theoretically predicted slope $2/5$. 
The values of the parameters are $\tau=1$ and $k=2.5$; the initial state 
is $\hat \psi_n(0) =\delta (0)$.
(b)  The logarithm of the coefficient 
of the sub-diffusion as a function of $\log(u)$, for the 
second moment (circles) and its time-average (squares). 
Empty and full symbols refer respectively to (M1) and (M2).
The dashed lines show the predicted dependence $\sim u^{4/5}$.}
\label{fig-loc}
\end{figure}

In this Section, we consider the regime 
where the value of the period of the $\delta$-kick perturbation is 
incommensurate with $4\pi$. For better comparison we fix $\tau=1$ for
all numerical computations shown in the following. 
The correspondent system in the $u\to 0$ limit is characterized by the 
phenomenon of dynamical localization \cite{chir}
caused by quantum interference effects.
Previous theoretical predictions and numerical simulations 
\cite{Shep1993,AR2003} 
indicate that, above a critical border $u_c\sim 2\pi$ for the  nonlinear 
coupling constant, dynamical localization is destroyed. The  
delocalization takes place in the form of 
anomalous subdiffusion with an exponent of $2/5$: in \cite{Shep1993} an 
asymptotic law $\langle p^2(t)\rangle\sim c(u)\cdot t^{2/5}$, 
(where $c(u)\sim u^{4/5}$) is predicted for both models; this 
is confirmed for both models by the data reported in Fig.~\ref{fig-loc}. 
In Fig.~\ref{fig-loc}(a) a bilogarithmic plot of the time-averaged 
second moment vs time is shown for increasing values of the 
nonlinear coupling constant $u$. In spite of large 
oscillations, both models fit the predicted asymptotic behavior with 
a power-law exponent equal to $2/5$.
For nonlinear coupling larger than the critical border $u_{c}$, 
marked by the vertical line, 
the dependence of $c(u)$ vs $u$ is confirmed for both 
models by Fig.~\ref{fig-loc}(b).
$log(c(u))$ is obtained by a one-parameter linear fitting of 
the logarithm of the second moment, once an anomalous diffusion 
exponent  equal to $2/5$ is assumed. 

\begin{figure}
\centerline{\epsfxsize=8.5cm \epsfbox{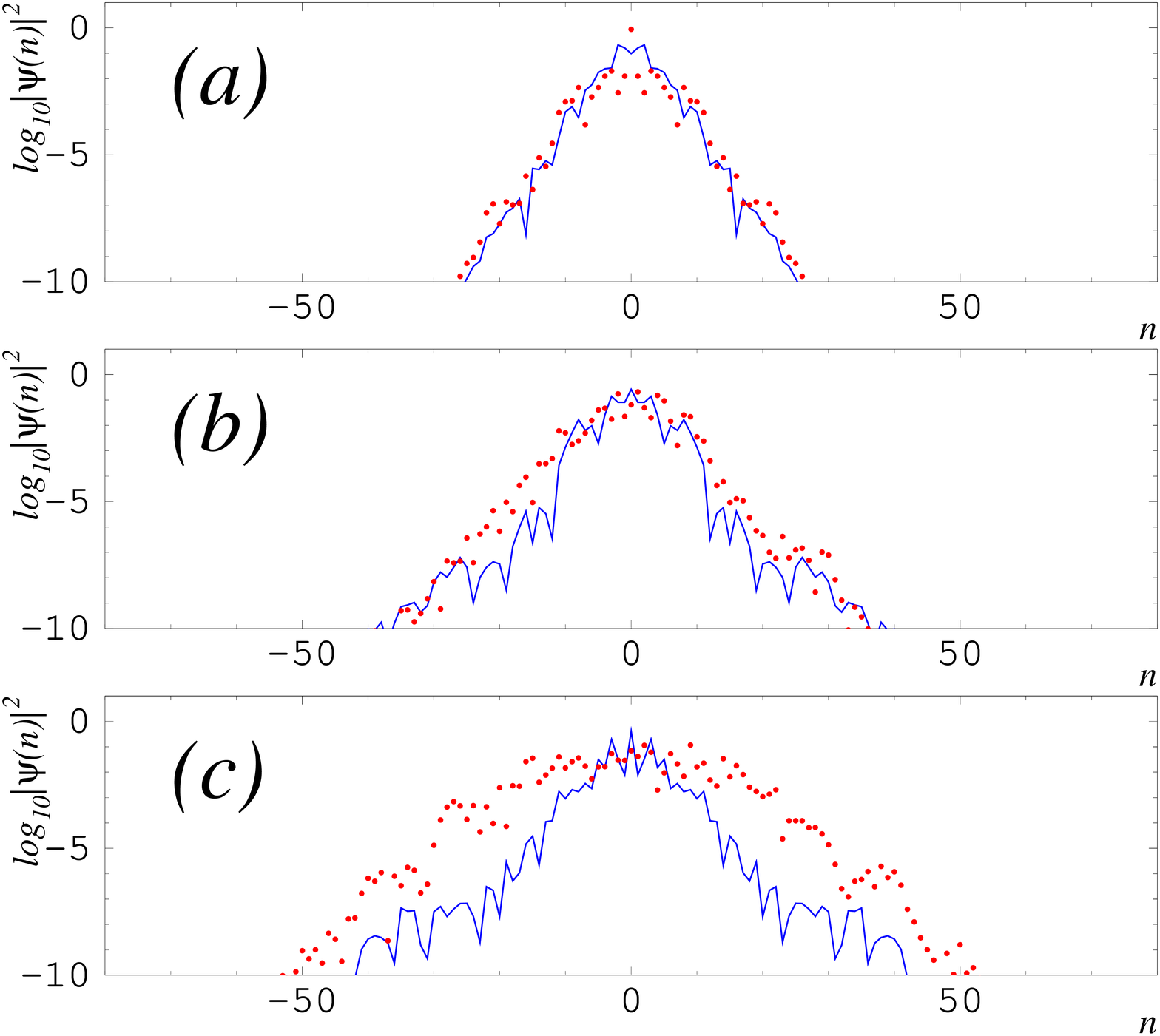}}
%\centerline{\epsfxsize=7cm \epsfbox{dipbeta-loc.eps}}
\caption{\small 
Comparison between the momentum distributions for (M1), circles, 
and (M2), full line, after $t=10$ (a), $t=100$ (b), $t=1000$ (c)
kicks. The parameters are the same as in Fig.~\ref{fig-loc}; the 
nonlinear coupling is fixed to the value $u=10$.} 
\label{fig-lst}
\end{figure}

The results obtained by calculating 
the evolution of the system with either (M1) or (M2), starting 
from the same state $\hat \psi_n(0) = \delta (0)$ 
and parameters, appear to be different on short time scales; nevertheless 
the two models share the same asymptotic behavior of the time evolution of 
the second moment and of the dependence of $c(u)$ 
vs $u$. The effect of the nonlinearity is the same for both models 
only at $t=1$, because of the common initial state. As explained in the 
introduction, the way nonlinearity acts on the wave function is essentially
different for (M1) and (M2), at least before the state becomes delocalized:
so deviations are qualitatively expected for intermediate times, 
while we expect a closer analogy in the models' behavior in the 
asymptotic regime.
Actually the close behavior exhibited by both models in Fig.~\ref{fig-loc} 
after a few time steps extends to more general features than the 
second moment: in Fig. ~\ref{fig-lst} we provide a comparison between full 
distributions over momentum states for $u=10$.

We remark that nonlinearity-induced delocalization has recently been 
explored also in studying survival probability on a finite 
momentum sample \cite{KW}: while the authors use (M2) to cope with 
computational difficulties, our findings suggest that their results
are probably relevant for true nonlinear Schr\"odinger equation dynamics too.

%%%%%%%%%%%%%%%%%%%%%
%%%%%%%%%%%%%%%%%%%%%

%%%%%%%%%%%%%%%%%%%%%
%%%%%%%%%%%%%%%%%%%%%

\section {Resonant regime}

%%%%%%%%%%%%%%%%%%%%%
\begin{figure}
\centerline{\epsfxsize=8.5cm \epsfbox{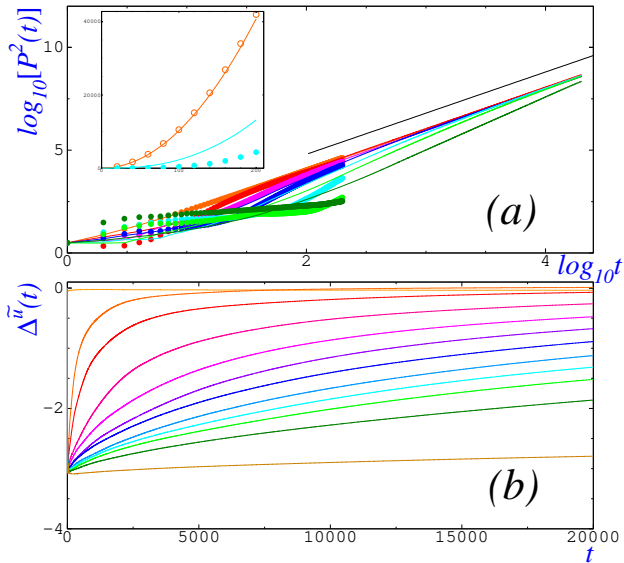}}
%\centerline{\epsfxsize=7cm \epsfbox{p2-res-delta.eps}}
\caption{
(Color online)
(a) Resonant growth of the time averaged 
second moment vs time in the 
presence of nonlinearity, for $\tau=4\pi$ and $k=2.5$. The initial
 momentum distribution is $\hat\psi_n (0)=\delta(0)$. 
The symbols and the full lines refer 
to (M2), with $N=2^{17}$, and (M1), with $N=2^{10}$ 
and $L=5000000$, respectively. 
The straight black line shows the resonant asymptotic behavior $t^2$. 
The values of the nonlinear parameter are 
$u =1,\ 5,\ 10,\ 20,\ 50,\ 100,\ 400$.
The inset is a magnification for $u=1$  and $u=50$ (lower part). 
A slight deviation between the two models can be seen for $u=50$.
(b) The function $\Delta^{(\uu )}(t)$ vs time for (M2).
Starting from above in the low $t<5000$ region, the values of the 
nonlinear parameter are $\uu= 0.1, 5, 10, 20, 30,
40, 50, 60, 70, 80, 100, 400$.}
%% saw ????
%%and the sub-critical value $\tilde u=0.5<\tilde u_{c}\simeq 1$. }
\label{fig-res}
\end{figure}
%%%%%%%%%%%%%%%%%%%%%

%%%%%%%%%%%%%%%%%%%%%
\begin{figure}
\centerline{\epsfxsize=8cm \epsfbox{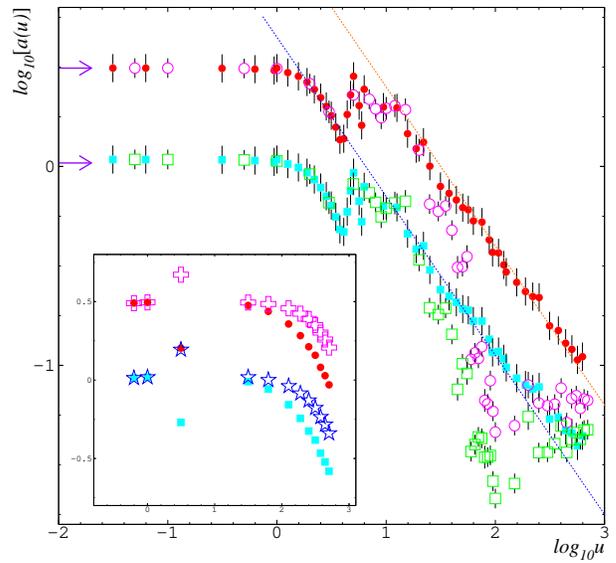}}
\caption { 
(Color online)
Bilogarithmic plot of the coefficient $a(u)$ of the quadratic 
growth of the second moment (circles) and its time-average (squares) 
as a function of the nonlinear parameter $u$.
Empty and full symbols refer to (M1) and (M2), respectively. 
The dashed 
lines show the algebraic behavior of $a(u)$ for large $u $ with an 
exponent equal to $-4/5$. Notice that for $u\leq 1$ the coefficients approach
the theoretical values of the KR-model, marked by arrows. In 
the inset the empty symbols refer to $a_\Delta$ calculated using 
the function $\Delta(t=t_{max})$ with $t_{max}=20000$.} 
\label{fig-coe}
\end{figure}
%%%%%%%%%%%%%

In this Section, we examine in detail the response of the system to
the nonlinear perturbation in 
the resonant regime of the KR ($\tau =4\pi r/q$ with $r,q$ relatively 
prime integers), characterized 
by a parabolic growth in time of the variance of 
the momentum distribution \cite{IzShep,Izr1990}.
The value of $\tau$ is chosen equal to $4\pi$, corresponding to the first 
fundamental quantum resonance of the KR.

In Fig.~\ref{fig-res}(a) a bilogarithmic plot of the time-averaged second 
moment of the momentum distribution for different values of the nonlinearity
is shown. The nonlinear coupling constant $u$ varies from 1 to 400. 
As already noticed \cite{AR2003}, the resonant behavior survives even in the 
presence of nonlinearity, although generically the spreading is slowed w.r.t.
the linear case.
On asymptotically long time scales, the resonant growth with quadratic 
exponent is reached even for strong nonlinear perturbations, though we observe that
the time needed 
to reach the asymptotic regime grows with $u$.  
The results in  Fig.~\ref{fig-res}(a)  
obtained from (M2), shown in full lines, 
allow us now a more detailed analysis of the behavior 
of the system on quite long times. It can also be seen 
in Fig.~\ref{fig-res}(a) that the time evolution of the second moment 
of (M1), shown with circle symbols, 
approaches the same asymptotic growth, even if some differences 
between the two models appear expecially 
for large nonlinearity ($u  \gtrsim 50$). \\

The persistence of the resonant behavior in the presence of nonlinearity 
can be explained intuitively as follows.
In the linear (i.e. $u=0$) resonant case the width of the momentum 
distribution increases linearly in time. Therefore, from the normalization 
condition, the probability amplitude to find the system in a momentum 
eigenstate $n$, decays as $|\hat \psi_n|^2
\sim 1/\Delta n\sim 1/(\pi kt)$ \cite{Shep1993,WGF2003}. 
The nonlinear phase shift 
$\tau u |\hat \psi_n|^2$ decreases with the same rate and its effects 
become irrelevant on long time scales, i.e. $t\gg \frac{u}{ \pi k}\tau$. 

Nevertheless, the nonlinearity affects the evolution of 
the second moment on smaller 
time scales $t\lesssim \frac{u}{ \pi k}\tau$ and introduces a $u$-dependence 
in the prefactor of the parabolic growth law, 
$\langle p^2(t)\rangle\sim a(u ) \cdot t^2$. 
Increasing nonlinear couplings manifests in a slower 
quadratic growth. 
In Fig.~\ref{fig-res}(b) the function $\Delta^{(\uu )} (t)
 =\left( \langle p^2(t)\rangle^{(\uu )} -\langle p^2(t)\rangle^{(0)}
\right)/t^2$ is plotted for different values of the nonlinearity. 
%This function, removing  the quadratic 
%dependence of the second moment, allows us to point out the different growth 
%rates for different values of $\uu$. 
Increasing asymptotic absolute values of $\Delta(t_{max})$ 
give an estimate of the modifications in the transport
induced by the nonlinearity.
The coefficients $a_\Delta$, calculated from the function 
$\Delta(t=t_{max})$ with $t_{max}=20000$ are 
shown in the inset of Fig.~\ref{fig-coe}.  

A detailed analysis of the dependence of the coefficient $a(u)$ of 
$\langle p^2(t)\rangle$ is presented in Fig.~\ref{fig-coe}. 
The numerical 
calculation of $\log(a(u))$ is obtained by a one-parameter linear fitting of
the logarithm of the second moment vs the logarithm of
 time with a straight line of fixed slope 2. The fitting is performed on a 
time interval $\Delta t = 200$; this rather small time interval was chosen 
in order to make a comparison between the results from
both models (empty and full symbols refer to 
(M1) and (M2), respectively). 
The accordance between the two models is satisfying up to $u \lesssim 50$. For 
$u>50$ the lowest order split method \cite{split} to evaluate  
the Floquet operator in (M1) becomes less stable and the numerical errors 
around the borders of the finite basis propagate faster.  

Numerical data are compatible with an 
algebraic law $a(u)\simeq {k^2}/(2(1+u/c)^\gamma$),
where $\gamma$ is $4/5$ and $c$ is a constant of the order of 10 (for the 
time-averaged moment the constant ${k^2}/{2}$ is substituted by ${k^2}/{6}$).
This law has the required asymptotic behavior for $u\to 0$: in this limit
$a(u)$ tends to the well-known value of the coefficient of the resonant KR, 
i.e. $a(0)={k^2}/{2}$ . In  Fig.~\ref{fig-coe} the values of $a(0)$ are marked 
by arrows.  For large values of $u$, $a(u)$ decreases for 
increasing nonlinearity with the inverse power law $\sim u^{-4/5}$. 
At the moment we have no explanation for the minimum observed in the 
intermediate region ($log(u)\sim 0.85$).

\begin{figure}[t]
\centerline{\psfig{figure=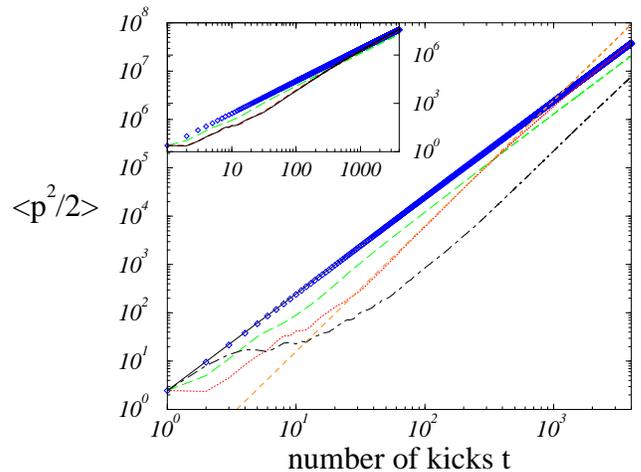,width=6.4cm,angle=270} }
\caption{Kinetic energy at the fundamental quantum resonance
$\tau =4\pi$, $k=\pi , p_{\mbox{\tiny \rm initial}}=0$, and
nonlinearities $\uu=0$ (solid), $-0.2$ (diamonds), $-1$ (dashed), 
$-10$ (dotted), and $-100$ (dash-dotted). The short dashed line
shows the superballistic increase $\propto t^{2.6}$ for the case $\uu=-10$.
The inset presents the results for $\uu =0.2, 1, 10$, 
and the case of $\uu=-10$ (thin solid line) for better comparison.
}
\label{fig2}
\end{figure}

\begin{figure}[t]
\centerline{\psfig{figure=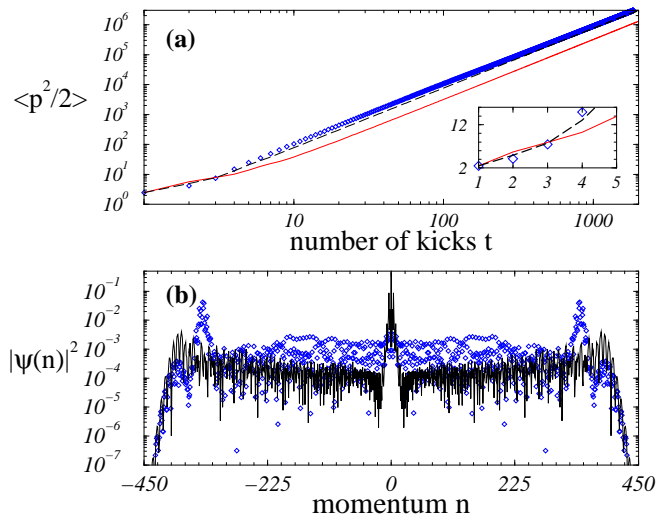,width=7cm,angle=270}}
\caption{(a) kinetic energy as a function of the number of kicks, and
(b) the corresponding momentum distributions after $200$ kicks,
for the quantum resonance
$\tau=3\pi$, same parameters as in Fig. \ref{fig2}, apart the nonlinearity
which is $\uu=-0.2$ (diamonds) and $0.2$ ((a) dotted, (b) solid). In (a) we show also
data for $u=0$ (dashed) for comparison, and the inset illustrates the 
opposite effect of the second and fourth kick (the rephasing in momentum space
occurs for the $u=0$ case only every second kick at $\tau =3\pi$) 
depending on the sign of $\uu$.
}
\label{fig3}
\end{figure}

%Looking closer again at the results presented in Fig.~\ref{fig-res}(a),
%we generally observe actually three regimes of resonant energy growth,
%which we characterize in the following revering to Fig.~\ref{fig2} for
%reasons of clarity:
Up to now, we discussed only the case of attractive interactions, i.e.
$u>0$. It turns out that the fundamental quantum resonance 
at $\tau =4\pi$ is insensitive to the sign of the nonlinearity as it can be
seen in the inset of Fig. \ref{fig2}. The same is true for the momentum
distributions which are not presented here. On the other hand, the next
order resonance at $\tau=3\pi$ is sensitive to the sign of $u$. For
$u<0$ in Fig. \ref{fig3}(b), the momentum distribution is slightly different
from $u >0$. Asymptotically, however,
the same ballistic growth of the mean square momentum is obtained.
This means that the details of the effect of nonlinearity depend on the 
resonance type as far as the sign of $u$ is concerned. This originates 
from the fact that while at the fundamental quantum resonances $\tau=4\pi m$
($m>0$ integer) the free evolution phase in the linear rotor is exactly one,
at higher order resonances there is a nontrivial phase evolution between two
successive kicks. The inset of Fig. \ref{fig3}(a) 
highlights that $\langle p^2\rangle /2$ either decreases
or increases with respect to the case $u=0$ at the second and fourth kick.
This is related to the fact that at $\tau=3\pi$ the rephasing in momentum
space occurs only every second kick, not between two successsive kicks 
as at $\tau=4\pi$.

In Figs. (\ref{fig2},\ref{fig3}) the time scales relevant for experiments
($t \lesssim 500$) are investigated (results refer to (M2)). In
Fig. (\ref{fig2}), we generically observe three regimes:
(i) there is an initial stage, where the mean square momentum increases
much slower than in the case $u=0$. This stage is followed by stage (ii)
where the increase can be faster than ballistic, and the mean square
momentum can even be larger for larger nonlinearity
(cf. $|\uu | = 10$ as compared to $|\uu |=1$ 
in Fig.~\ref{fig2}). \\
The observed superballistic growth of the second moment of the
momentum distribution is quite surprising, in particular, having in mind that
such a growth is forbidden in the usual KR (i.e. $u=0$) \cite{casati1986}.
The results in Fig.~\ref{fig2} are reminiscent of the observed superballistic
spreading in 1D tight-binding models \cite{ketz2001}, however, here
the superballistic behavior is caused by the {\em nonlinear} term in the
time evolution, in contrast to the linear Hamiltonian models in 
\cite{ketz2001}. 
In terms of the model studied in \cite{ketz2001}, the nonlinearity $u$ 
would act as finite size trapping region (cf. also \cite{KW}), outside 
of which the motion is ballistic (we already showed how nonlinearity 
does not essentially modify high $n$ components).
   
The final stage
(iii) we call asymptotic regime, because there the growth exponent
approaches the 
one for vanishing nonlinearity (only for $\uu = -100$ this stage is not yet
reached in Fig.~\ref{fig2}). 

\begin{figure}
\centerline{\psfig{figure=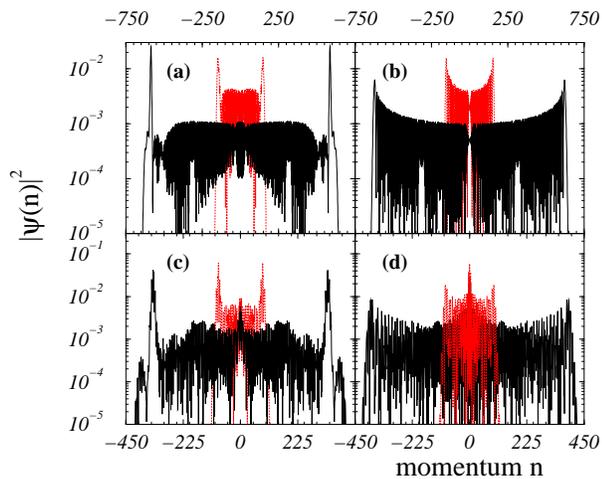,width=6.5cm,angle=270}}
\caption{Momentum distributions for zero initial momentum, 
$k=\pi$, $\uu=-0.2$ (a,c) and $u=0$ (b,d), and 
$\tau=4\pi$ (a,b), $\tau=3\pi$ (c,d). Note the stable
peaks at the largest momenta in the case with small nonlinearity. 
The distribution are shown after 50 (dotted) 
and 200 kicks (solid line) in each panel.} 
\label{fig1}
\end{figure}

\begin{figure}
\centerline{\psfig{figure=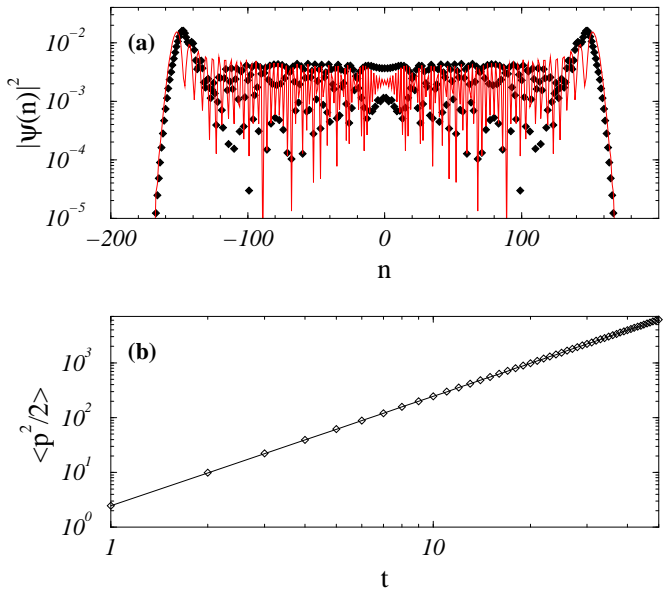,width=8cm,angle=270}}
\caption{
(a) momentum distribution after $t=50$ kicks, and (b) mean square momentum
for $p_{\mbox \tiny \rm initial}=0$, 
$u =-0.2$ and $\tau=4\pi, k=\pi$, shown are the evolutions induced
by (M2) as diamonds, and by (M1) as full lines, respectively.}
\label{fig4b}
\end{figure}

In Fig.~\ref{fig3}(b) 
we notice the distinct peaks close to the very edge of the
momentum distribution, for $\uu = 0.2$. Such peaks have been found for
sufficiently large kicking strength $k \gtrsim 2.5$ and corresponding
$|\uu (k)|= 0.2\ldots 2$, and it turns out that they can be up to one order of
magnitude higher that the maximum of the momentum distribution for
the linear KR at the resonances $\tau = 4\pi$ and $\tau = 3\pi$.
Fig. \ref{fig1} compares the momentum distribution
$|\hat\psi_n|^2$ at the fundamental quantum resonance $\tau=4\pi$, and at the
resonance $\tau =3\pi$ for small nonlinearity $\uu =-0.2$
with the case of the linear KR. 
The distributions are shown after 50 and 200
kicks, respectively, to stress their evolution in time. For both
resonances, we observe a very interesting feature, namely, the small
nonlinearity 
sharpens the edge peaks, which
%enhances the probability to be at a particular momentum range, 
%and the corresponding peaks
move ballistically, i.e. with a speed which is
proportional to the number of kicks $t$
(we recall that when $u=0$ the distribution is characterized by a largest momentum
component also moving according to a linear law $n_{max}(t)\simeq kt \pi/2$ \cite{WGF2003}). 
The peaks are more pronounced than
%any probability 
in the linear case, and are remarkably stable, i.e. their height
decreases very slowly with increasing number of kicks in Fig. \ref{fig1}(c),
or even increases initially as in Fig. \ref{fig1}(a).
While we focussed our discussion on the model (M2), the structure of the probability
distribution is quite similar for (M1), see Fig. (\ref{fig4b}).
%At large momenta the nonlinearity
%has little effect, because $|\hat\psi_n|$ is practically zero outside the 
%largest momentum component. The latter moves in time, and it is given
%in the case of $u=0$ by $n_{\mbox{\tiny max}}(t)\simeq kt\pi/2$ \cite{WGF2003}.
%For $n<n_{\mbox{\tiny max}}$, 
%the nonlinearity strength is given by the local weight
%of $|\hat\psi_n|^2$, which as the momentum spreads to larger values decreases
%everywhere except at the enhanced peaks. The latter come about
%by the interplay of the nonlinearity effects and tendency
%of ballistic spreading due to exact phase revival in-between two kicks, which
%can occur only if the nonlinearity is negligible, i.e. 
%for $n\sim n_{\mbox{\tiny max}}$.

%As discussed in the introduction, we have to keep in mind that our model
%(M2) should be compared with the
%type of nonlinearity which characterizes a Bose-Einstein condensate.
%If its motion is defined on a ring, it can be modeled by the evolution
%(M1). We have computed this evolution for the same parameters
%in the case of $\tau=4\pi$, and have found very close agreement between the
%two cases. We have used 100000 Trotter-Kato steps in the discretization
%of the ``free''  evolution part of (M1). 
%The result in Fig. \ref{fig4b} shows that both models based
%either on (M1) or (M2) lead to almost the same dynamical
%behavior. This confirms again what was found above.

The intermediate time scaling properties look 
in this case more complex than in the kicked rotator dynamics in 
the presence of sticking accelerator modes, where the same exponent
appears both in the classical and in the quantum case (where a new 
modulation appears). Work is in progress to see whether there 
exist classical mappings that reproduce the peak dynamics we 
observe in intermediate time quantum behavior \cite{allegrini1998}.

\section {Momentum diffusion in the resonant regime}

All the above results have been obtained for 
an initial state in the form of a plane wave of null momentum.
%constant initial momentum
%$p=\beta=0$. 
In a typical experiment, one can create a Bose condensate with
an initial spread of momentum which is much less than two photon recoils
(which can be imparted as momentum kicks to the atoms by the kicking laser).
We ask ourselves what happens if such a spread is taken into account. 
The momentum variable $p$, varing on a discrete lattice in
the case of single rotor, becomes a continuous variable. 
For the linear KR, the eigenvalues of $\hat p$ can be written, 
distinguishing the integer and fractional part (quasi-momentum), 
as $p=[p]+\{p\}=n+\beta$. Owing to the conservation of quasi-momentum $\beta$,
the system dynamics can be decomposed 
in a bundle of rotors \cite{FGR2003}
(called in the following $\beta$-rotors), each 
parametrized by a value of the quasi-momentum, evolving incoherently 
with operators with the same functional form of 
(\ref{model1}) and  (\ref{model2}), in 
which $\hat n$ is substituted by $\hat n+\hat\beta$.
Such a decomposition is not easily accomplished when we introduce a nonlinear term
in the dynamics. The general task we have to face becomes the study of a
nonlinear evolution equation with periodic coefficients. This is a quite a
complex problem that cannot be tackled in full generality, even though different
approximation schemes have been proposed, {\em e.g.} by mapping the problem into a 
discrete lattice, which turns out to be useful if the wave function is expanded in
a suitable set of localized functions related to the linear problem \cite{AKKS}.
We generalize (M2) in such a way that its linear limit is the evolution operator
corresponding to a quasimomentum $\beta$ (as formerly specified), and {\em assume}
that each {\em nonlinear $\beta$ rotor} evolves independently. In this way we study
the influence of nonlinearity on {\em realistic} initial conditions in a highly
simplified way, by means of a {\em generalized} (M2) model: further work is obviously
needed to check whether our findings extend to a full GP dynamics. 
%Since both of our models (M1) and (M2) preserve the spatial periodicity
%of the linear KR, we start from an assumption valid in the 
%limits of small nonlinear couplings and long times. We assume 
%that, even in the presence of a nonlinearity,  owing to the invariance 
%of the evolution operator $\Uop$ under spatial translations 
%by multiples of $2\pi$, the Bloch theory and the 
%consequent quasi-momentum conservation holds. 
%The evolution of the BEC is supposed to be correctly modeled by 
%an ensemble of atoms, whose dynamics is controlled by 
%independent nonlinear Schr\"odinger equations. 

The quantum resonance phenomenon in the KR is strongly sensitive to the values 
of the parameters of the system.  The linear KR rotor ($u=0$) 
exhibits the quantum resonance only for a finite set of 
quasi-momenta, i.e. $\beta= \beta^R =m/2p$ with $m<2p$ 
\cite{Izr1990,FGR2003,WGF2003}.
A slight deviation of the quasi-momentum from $\beta^R$
changes completely the evolution of the system.
For values of $\beta\neq\beta^R$, after a transient  
regime, the suppression of the  resonant 
growth of the energy of the linear $\beta$-rotor through 
the dynamical localization occurs; at fixed time $t$, 
only quasi-momenta within an interval $\sim 1/t$ of $\beta^R$ mimic 
the ballistic behavior $\propto t^2$ and a rough estimate 
of the time up to which the quadratic growth of the $\beta$-rotor 
energy persists is $\bar t\sim 1/\Delta \beta$, 
where $\Delta\beta =|\beta-\beta^R|$ \cite{WGF2003,WGF2004}.

%%%%%%%%%%%%%%%%%%%%%
\begin{figure}
\centerline{\epsfxsize=8.5cm \epsfbox{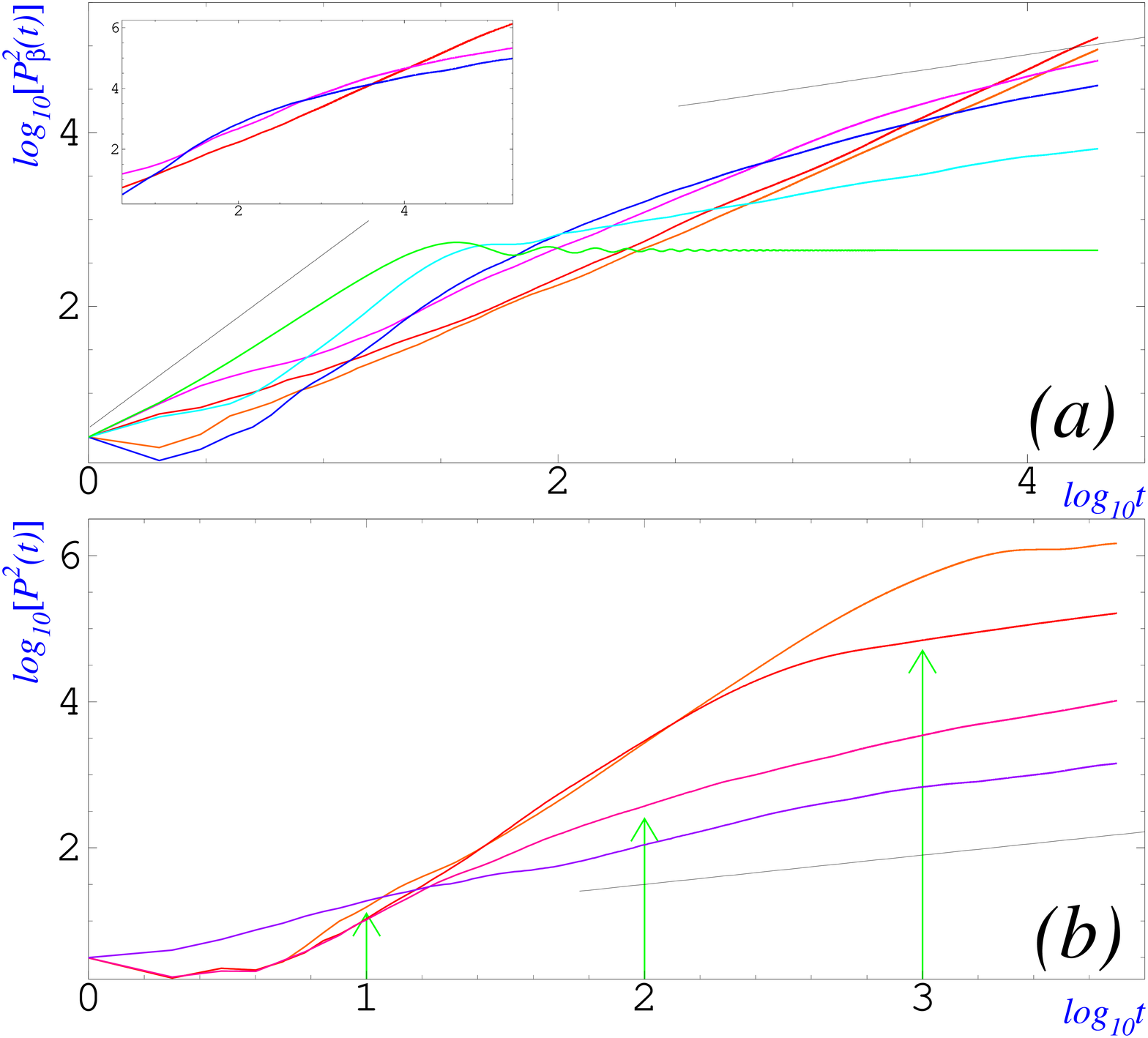}}
\caption{\small 
(Color online)
(a) Bilogarithmic plot of the 
second moment of a single $\beta$-rotor with 
a fixed quasi-momentum for increasing nonlinear coupling (from below,
referring to high t values,  
$\uu =0,1,5,10,50,100$). The two lines have slope $2$ and $2/5$. Note 
that in the KR case ($u=0$), the localization occurs for time $t
\gtrsim 1/\beta$. The parameters are $\tau =4\pi$, $k=2.5$ and $N=2^{17}$. 
The initial state 
is an eigenfunction of the momentum with $n=0$ and $\beta=0.00946556$. 
In the inset the calculations are prolonged ten times. The
unbounded growth for $\uu=50,100$ can be clearly seen.
(b) The same as (a) with $u$ fixed ($\uu=5$) and variable $\beta$ (from 
above, referring to high t values,
 $\beta =0.0001, 0.001, 0.01$ and $0.1$). The arrows mark the times 
$1/ \beta$.} 
\label{fig-1qm}
\end{figure}
%%%%%%%%%%%%%%%%%%%%%

%First of all we consider the effects of the nonlinearity on the 
%dynamics of a single $\beta$-rotor with fixed 
%quasi-momentum. 
In the following, we investigate 
%in greater detail
the mean square momentum distribution of the generalized (M2) model, 
first for a) for a single $\beta$ rotor with fixed quasi-momentum, and
then b) for an incoherent ensemble of $\beta$-rotors 
whose initial state in momentum space is a Gaussian distribution with zero mean
and rms spreading $\Delta \beta = \sigma =0.01$.
%as done in Fig. 
%\ref{fig4} above. 
We choose $\tau=4\pi$,
and the resonance condition is then met for $\beta^R =0$ and $\beta^R =1/2$.
The case with $\beta^R =0$ fixed was considered in Section III.
As in the localized regime, considered in 
Section II, the introduction of the nonlinearity causes a delocalization 
in the system with a non-resonant value of the quasi-momentum
($\beta\neq\beta^R$).
For small nonlinearities, the appearence of an anomalous asymptotic  
diffusion with an exponent  
of $2/5$, after the initial ballistic behavior, 
is confirmed by data of Fig.~\ref{fig-1qm}(a) for a $\beta$-rotor with $\beta
\approx 0.009$. On the contrary, greater nonlinear 
couplings ($u\geq 50$) introduce 
an excitation of diffusive type, starting from the first kicks.
In Fig.~\ref{fig-1qm}(b) the quasi-momentum of 
the $\beta$-rotor is varied and $u$ is kept fixed. 
The arrows mark the times $\bar t$, depending on the value of $\beta$,
approximately bounding the region of the ballistic growth.

\begin{figure}
\centerline{\epsfxsize=8.5cm \epsfbox{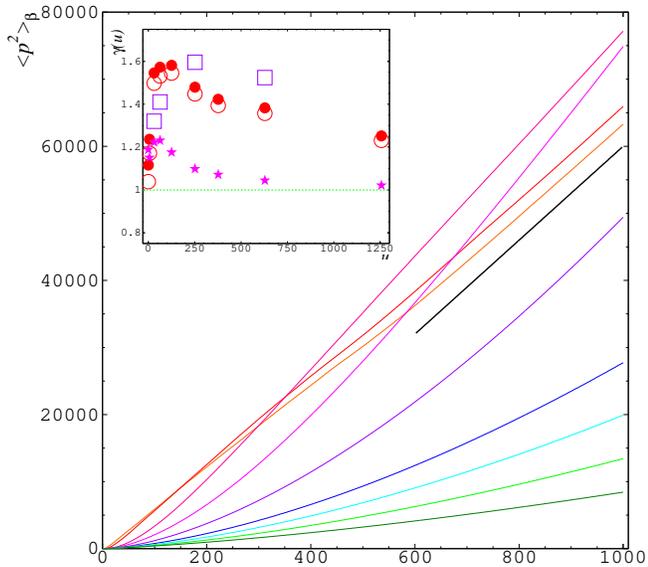}}
\caption{\small
(Color online)
Average over 5000 quasi-momenta of the second moment of the distribution 
vs time for (M2); $\tau = 4\pi$ and $k=2.5$. Different values 
of the nonlinear parameter are shown in different colours; starting from 
below (referring to high t values) $\uu = 200,\ 100,\ 60,
\ 40,\ 20,\ 0,\ 1,\ 5,\ 10$; 
the initial wave packet in Fourier space is a 
Gaussian distribution centred in $n=0$ with rms $\sigma =0.01$. 
The inset shows the power-law exponents of the second moment as 
a function of $\uu$. The fitting is performed on time intervals 
$\Delta t=1000$ (circles)
 and $\Delta t=6000$ (squares). Open symbols refer to 500 quasi-momenta.
Stars refer to a uniform distribution of quasimomenta.}
\label{fig-ave}
\end{figure}

We then consider the dynamics of an incoherent ensemble 
of $\beta$-rotors.
%Since it is assumed that no interference occurs between 
%different $\beta$-rotors, the expectation values of 
%the physical quantities are obtained by a superposition of those of the 
%independently evolving rotors. 
The mean square displacement of the distribution is 
$\langle p^2(t)\rangle_\beta =\int d\beta\langle p^2_\beta(t)\rangle$. 
The average over $\beta$ has been calculated using $5000$ quasi-momenta.
In Fig.~\ref{fig-ave} the time evolution 
of the averaged second moment of the initially 
Gaussian wave packet is shown for (M2). The behavior in the 
correspondent linear case of the KR is theoretically known
(see Appendix A of \cite{FGR2003}): for $u=0$, the kinetic energy of the system increases 
diffusively in time with a coefficient 
proportional to $k^2/4$, and dependent on the initial distribution of 
quasi-momenta \cite{WGF2003}. 
The presence of the nonlinearity manifests itself in a faster 
than linear growth, at least on short time intervals. After this transient regime, 
the asymptotic growth is expected to become approximately linear.
The black straight line is drawn for better comparison.
At fixed time $t$ and assuming an uniform distribution of the
quasi-momenta, the resonant rotors, whose quasi-momenta lie within 
the interval $\Delta\beta$, enter in the average of 
$\langle p^2\rangle_\beta$ with a contribution of $\sim t^\alpha$ 
and a weigth $w\sim 1/t$, while 
the non-resonant rotors give a contribution of $(1-w)*t^{\gamma}$.
The exponents of the transport in the limit $t\to +\infty$ 
reach the values $\alpha (\infty)=2$ and $2/5\leq\gamma \leq1$. 
Therefore, asymptotically in time, the global transport 
exponent reaches the value 1. 
In the inset of  Fig.~\ref{fig-ave} the exponents 
of the algebraic growth of the second moment are plotted as a function 
of the nonlinear coupling constant $\uu$. 
The fitting time intervals is $1000$ kicks. 
Full and open circles refer to 
5000 and 500 quasi-momenta of the initial Gaussian distribution:
a slight rise in the 
exponents can be noted increasing the number of quasi-momenta, because a 
greater amount between them approches the value $\beta^R=0$, yielding the 
quadratic growth of the $\beta$-rotor energy.
Note that the exponent approches faster the value $1$ for a uniform 
initial distribution of quasi-momenta (stars), confirming the previous 
argument.

\begin{figure}
\centerline{\psfig{figure=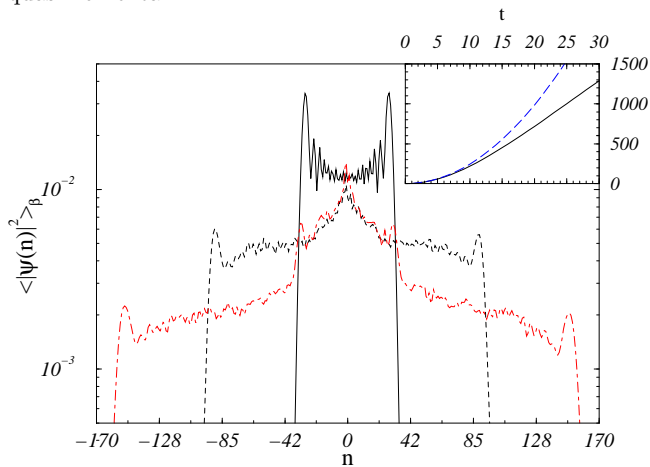,width=6cm,angle=270}}
\caption{Momentum distribution after $t=10$ (solid), $30$ (dashed), and
$50$ kicks (dash-dotted), for the same parameters as in Fig. \ref{fig1}(a), 
but incoherently averaged
over independently evolved initial conditions (Gaussian initial 
momentum distribution with rms $\sigma =0.01$ centred around $n=0$). 
The inset shows the corresponding $\langle p^2 \rangle _\beta /2$
as a function of the number of kicks for $\uu =-0.2$ (solid) 
and $u=0$ (dashed).
}
\label{fig4}
\end{figure}

Fig. \ref{fig4} presents a closer look at dynamics on a shorter time scale:
the results refer to the case in which we found
stable momentum peaks in Fig. \ref{fig1}. Part of the peak is still preserved
for the used spread $\Delta \beta \simeq 0.01$, which can be realized in  
state-of-the-art experiments \cite{expintro,Pisa2001}. After about 15 kicks, 
more weight lies, however, now in the centre of the distribution made up 
of rotors which do not exactly fulfill rephasing condition due to 
nonzero quasimomenta. Also the increase
of the mean square momentum, which is averaged incoherently over all 
the independently evolved initial conditions, is then 
not any more quadratic but closer to linear (see inset in Fig. \ref{fig4}),
as was found in the case of a uniform initial distribution
of quasimomenta for the $u=0$ case \cite{WGF2003,WGF2004}. 
The mean square momentum
still increases much faster than for nonresonant values of the kicking period
$\tau$, where dynamical localization occurs. The latter may be destroyed
by the nonlinearity but the above observed growth of $\langle
p^2\rangle /2 \propto t^{2/5}$ (cf. Section II)
is much slower than linear. 
On short time scales thus quantum resonance is very robust with respect to nonlinear 
perturbations. If our incoherent superposition model is correct after 
some initial stage, the ballistic motion should cease but the dynamics
will show the influence of the ballistic quantum resonant transport.

%%%%%%%%%%%%%%%%%%%%%
%%%%%%%%%%%%%

\section{Conclusions}
\label{concl}

In summary, we have numerically analysed in great detail the quantum transport
occuring in two nonlinear generalizations of the
famous $\delta$-kicked rotor model, with a cubic nonlinearity as present in
the Gross-Pitaevskii equation. We confirm previous results in the
regime of localized transport, and show the validity of the predictions
of \cite{Shep1993} for a wide range of nonlinear coupling strengths. 
In addition, we found that the quantum resonances of the kicked rotor are
very stable with respect to the nonlinear phase perturbation, which
looses its effect in the asymptotic limit of large interaction times
with the periodic driving. Surprising phenomena like pronounced peaks
in the momentum distributions at quantum resonance, and superballistic
intermittend growth of the mean square momentum have been found. Both 
phenomena are caused by the cubic nonlinearity in the evolution, which shows
that the analysed models bear a rich dynamical behavior in parameter space. 

The experimental works on the kicked rotor using a Bose-Einstein condensate
\cite{expintro} mostly concentrated on the
short time behavior at quantum resonance or on the socalled anti-resonance,
where the motion is exactly periodic in the case $u=0$.
But an experimental observation of the ballistic quantum resonance dynamics
up to $10\ldots30$ kicks seems possible, for small enough kicking strength $k$
such as to avoid a too fast spread in momentum space which cannot be monitored
by standard time-of-flight detection \cite{WGF2003,darcy2004}.
Our results are fully consistent with the few published experimental data, 
which show that both resonance and antiresonant dynamics essentially 
survive the presents of small nonlinearities, apart from other effects which
damp, for instance, the periodic oscillations at the anti-resonance. Such
effects are e.g. 
the uncertainty of the centre of the initial momentum distribution,
and fluctuations in the experimental kicking strength \cite{expintro}.

%%%%%%%%%%%%%%% \acknowledgments %%%%%%%%%%%%%%%%%%%%%%%%%%%%%%
LR and RA acknowledge partial support from the MIUR--PRIN 2003 project {\em 
Order and chaos in nonlinear extended systems: coherent structures, weak 
stochasticity and anomalous transport},  and the INFM Advanced Project\
{\em Weak chaos: theory and applications}.
SW thanks Prof. Ken Taylor for his hospitality and financial support
at the Queen's University of Belfast, where part of the present work 
originated. Enlightening discussions with Prof. E. Arimondo and
O. Morsch on the experimental possiblities, and with
R. Mannella on L\'evy statistics are gratefully acknowledged.

%%%%%%%%%%%%%%%%%%%%%%%%%%%%%%%%%%%%%%%%%%%%%%%%%%%%%%%%%%%%%%

\end{document}